\documentclass{ws-book9x6}
\usepackage{wrapfig} 
\usepackage{amsfonts} 
\usepackage{amssymb} 
\usepackage{amsthm} 
\usepackage{amsmath} 
\usepackage{rotating} 
\usepackage{braket} 
\usepackage{qcircuit} 
\usepackage{verbatim} 
\usepackage{listings}
\usepackage{mathtools}
\usepackage{ws-book-har}   
\usepackage[pdfpagelabels=false,colorlinks=true,allcolors=black]{hyperref}  

\title{Advancements in Quantum Computer Music}      

\makeindex

\begin{document}










\setcounter{page}{1}

\setcounter{chapter}{6}

\chapter[Qubit Instrumentation of Entanglement]{Qubit Instrumentation \\ of Entanglement}\label{ch-mc}

\author{Mark Carney}
  \address{Quantum Village Inc. \\ London, UK \& Delaware, USA\\ Email:
    mark@quantumvillage.org}

{\bf Abstract:}
This chapter and the experiments described within explore how `human entanglement' might be represented and even emulated by physical entanglement. To achieve this, a notion of `tonal centrality' between two musicians is captured via MIDI and passed as a parameter into a quantum simulation taking place on an embedded device (a Raspberry Pi Pico). The results of these simulations are then coded back into MIDI and sent to the players' instruments. The closer the musicians' tonality is, the more their instruments will be entangled in a $\ket{\Phi^+}$ state, and the further away they are the more their instruments will be entangled in a $\ket{\Psi^+}$ state. The intention is to create random parameters that are correlative - \emph{i.e.} the same on both instruments - or anti-correlative - \emph{i.e.} the bit-wise opposite of each other, influenced by the tonal relationship from the players. These random parameters sharing these particular properties add a new dimension for quantum-musical expression. This concept was realised experimentally, and the full code and sample outputs are provided. This work aims to pave the way for musicians to explore and experience quantum emulations of their own musical experiences, adding a new nuance and possibilities for the future of \emph{entangled ensembles.}
\newpage

\section{Introduction}

Quantum information theory has fast become the dominant model within which we express algorithms that involve quantum computing and quantum information processing \cite{Nielsen_Chuang_2010}. Alongside the general paradigm of quantum information science and technology (QIST), the quantum arts has started to emerge with quantum computer music, in many ways, leading the charge - see for many examples the pioneering tome `Quantum Computer Music' edited by Miranda in \cite{Miranda_et_al_2022} as well as \cite{Miranda-2022}. The main artistic exploration encapsulating this work may be stated as follows:

\textbf{Core Idea} - Can metaphysical entanglement be made physical? To what end may two instruments react in an entangled way to the `musical entanglement' of the players?

The core quest for this author's work concerns methods for the `instrumentation' of the fundamental building block of QIST; the qubit. Seeking first to understand how one might `instrument' (as one would a tool) a valid qubit, and with such tooling this work then seeks to understand the qubit as an instrument itself; how do you find ways to create music from algorithms acting on qubits in response to input from musicians?

The qubit is a gateway to the special properties of quantum systems that are used to great interest - namely, superposition and entanglement \cite{Nielsen_Chuang_2010}. This chapter will discuss some very low-level implementations of these phenomena in software simulated environments that are kept at the lowest software level possible, that of embedded firmware on a microcontroller. The primary driver for the work presented in this chapter is giving a different view on how we can create music, as sculpted time, by shaping quantum/-inspired processed around such sonification.

A side technical note - this work does not find reason to distinguish between `real' hardware or `simulated' software qubits, mostly owing to the lack of any magic state preparation nor presence of any $T$ gates. 

This chapter will log the author's continuing progress in this particular quantum musical exploration. This section has seven sections:

\begin{enumerate}
    \item Section \ref{sec:bg} describes a walking tour of the musical and philosophical backdrop to considering why entanglement is a key area of exploration for quantum music. 
    \item Section \ref{sec:qit} describes the background quantum information theory - this is mostly illustrative and not essential.
    \item Sections \ref{sec:methodology} and \ref{sec:tooling} describe the methodology and experimental setup for creating embedded quantum simulations that allow the `entanglement' of two instruments. 
    \item Sections \ref{sec:results} and \ref{sec:results} detail the results of this work, and the last section \ref{sec:future} details exciting new directions this work may open up to experimentalists and instrumentalists alike.
\end{enumerate}

\section{Background and Motivation}\label{sec:bg}

The first paper from this author on quantum music may be found in \cite{Carney_2023}. In this work, the author created several options for utilising a simple quantum circuit as a `quantum effects unit', primarily intended to show that quantum simulations could be carried out at speed, on low power devices, and in a way that could be much more reactive to a musician's needs whilst preserving the player's access to the traditional instruments that that are perhaps much more familiar with. 

The primary motivation for extending the usage of embedded systems is to reduce the latency and jitter when interacting with the quantum phenomena harnessed to your instrument. Whilst the author's previous work in \cite{Carney_2023} furtively played with entanglement, it was left as a note of future work whether it would be practical to simulate entanglement between two instruments from a single simulated source. 

This chapter presents the success of harnessing the entanglement between musicians to influence the entanglement between two instruments, deepening the effect on a performance. First, however, it is important to state the background ideas, facts, and theories that drive the core intuition for the proposed solutions. 

This section will discuss some of the background work that underpins the activities and design decisions the author has made to create the proof of concept entangled instruments detailed later. The main drivers for this work are the desire to reduce the latency by embedding quantum interfaces in instruments, and to begin artistic explorations of entanglement that take their crystallisation point from entanglement between musicians that have already been observed at a higher level. 

\subsection{Latency in Musical Interfaces}

Latency and Jitter are considered the most salient timing properties when designing or building electronic instruments \cite{McPherson_2020}. Latency is commonly defined as the delay between an action and the intended response, \emph{e.g.} between pressing a piano key and hearing a note. Jitter is the variance in this response time, and for music is an important factor. If a system has a high latency, the player can sometimes adjust to compensate for it but only if the jitter is low enough as to make such a compensation reliable. 

In a paper on performance and instrument latency, \cite{Wessel-2002}, Wessel and Wright proposed that the optimal latency for music is around 10ms, whilst Fujii, Hirashima, Kudo, Ohtsuki, Nakamura, and Oda in \cite{Fujii_2011} determines that the target jitter should be around 1ms - roughly one quarter the jitter observed by human professional percussionists, who can reduce their manual jitter to just 4ms when playing a regular rhythm (cf. \cite{Fujii_2011}). 

These are incredibly tight tolerances, but as McPherson, Jack, and Moro demonstrate in \cite{McPherson_2020} such timings are easily attainable with modern microcontrollers and consumer/hobbyist electronics. Indeed, they were able to obtain 500$\mu s$ latency with 23$\mu s$ jitter by removing the USB-based device-computer interface, instead opting to embed the processing next to the core electronics responsible for handling the intended musical interactions. In fact, they point out that ``Even under a reductive best-case scenario, a microcontroller connected to a computer by USB fails to meet the 10ms action-to-sound latency standard.''\cite{McPherson_2020}

This background work, from a larger body of work on musical interface analytics and design, form the basis for why this work undertakes the task of embedding the quantum simulation and interface processing in the same device; to minimise both latency and jitter with the aim to make a wholly more fluid user experience for the exploration of quantum in music.

The author's previous work in \cite{Carney_2022} and \cite{Carney_2023} demonstrated that the use of embedded systems can be used to run quantum simulations with minimal latency and jitter. Despite there being some much more powerful platforms than the ones detailed in section \ref{sec:tooling}, such as the BELA platform first described in \cite{Mcpherson2015}, the prior existence of both code and electronics from the author's previous work determined the technical direction of the setup described below. That said, the classical nature of the simulations means that these experiments may be re-tooled for other platforms with minimal effort.

\subsection{Entanglement - a View from Philosophy and HCI}

The main focus of this work is to describe and interlink physical and musical entanglement. To that end, this section will summarize the background `entanglement theories' that this work looks to build upon.

Since the now infamous EPR paper \cite{Einstein_1935}, entanglement has posed many interesting questions. The physical definition of entanglement is usually along the following lines, here taken from \cite[p.95]{Nielsen_Chuang_2010}:

\begin{quote}
    ``We say that a state of a composite system having this property - that it cannot be written as a product of states of its component systems - is an \emph{entangled state}.''
\end{quote}

Entanglement is more than simply `two particles that are somehow linked'. This is a very profound statement about the deep mathematical nature of two individual particles that necessitates that their physical descriptions cannot be separated. This is anathema to the traditional view that dictates that particles are wholly separate entities from one another.

That is not to say that they are necessarily indistinguishable, where they cannot be told apart. The symmetrization that forms entanglement requires more nuance, such as Kaster attempts to create with their notion of ``Quantum Haecceity'' in \cite{Kastner2023}. A thorough philosophical analysis of quantum physical notions of indistinguishability, individuality, and how these relate to entanglement may be found in Catren's work \cite{Catren2023}. 

The mathematical concept and formulation of entanglement has lead to a surprising ontological interpretations since the inception of quantum mechanics in the early 20$^{th}$ century. Karen Barad explained the concept of `entanglement' in \cite{Barad_2007} as:

\begin{quote}
    ``To be entangled is not simply to be intertwined with another, as in the joining of separate entities, but to lack an independent, self-contained existence. Existence is not an individual affair.''
\end{quote}

The philosophical changes that are wrought by the existence of entanglement defy much of the `classical' understanding of the world that we have been forced to abandon in quantum mechanics. Such a vastly different interpretation of how things come `to be' - namely, that `what is' is comes into being from complex processes of both interaction and intra-action (see \cite{Zanotti2017}, \cite{Catren2023}). As Wendt points out in \cite{Wendt2015}:

\begin{quote}
    `` The fact that the parts of entangled quantum systems have only relational properties upsets the fundamental principle of atomism that the nature of wholes is determined by the attributes of parts, and even raises doubts about whether `parts' exist at all.''
\end{quote}

`Entanglement Theories' have developed and grown around the interactions between humans and the material world. There are many ways in which humans seem to be `caught up' in the existence of things in ways that are more than merely superficial. Frauenberger in \cite{Frauenberger2019} states:

\begin{quote}
    ``[Entanglement theories] all make a radical proposition: humans and things are ‘ontologically inseparable from the start’...  Entanglement theories [question] the locus of agency, asking which active contributions tools make to what humans do.''
\end{quote}

With the apparent entanglement of humans and their tools, it is natural to ask how does this extend to highly expressive tools such as musical instruments? This author's conjecture is that such deep human-object entanglement enhances and enables deep human-to-human entanglement of the kind seen in musical performances. A summary of findings which seem to confirm this is found in \cite{Bhave2023}, and is discussed in section \ref{sec:jazzentanglement}.

Some modern theories of ontology, such as Actor-Network Theory (ANT) have deep problems with such human-object entanglements owing to the complexities of two agents sharing the same higher-abstract description \cite[p.113]{Harman2018}. This difficulty arises from a primary tenet of Object Oriented Ontology (OOO) that dictates that all objects are distinct from each other \cite{Harman2018} - something that the existence of entanglement challenges at the most fundamental level. 

However, such challenges are not new to OOO; indeed Wolfendale in \cite{Wolfendale2014} has given an incredibly thorough, lengthy, and comprehensive analysis of OOO, settling on the critique that the claims by OOO to offer a non-correlationist viewpoint is something it fundamentally fails to deliver upon. Given this, it should not be unexpected that the notion of `entanglement' as defined above remonstrates such an approach. 

This creates a nuance and complexity about the status of existence of `mereological dynamic systems'; the author's own terms for a system that has its discourse principally around descriptions of its parts/smaller agents and their relationship to the greater whole or the networks that bind them (mereological), and that these systems evolve in time (dynamic). This work relies on the observation that  music is such a mereological dynamic system. 

How can we describe music that is necessarily constituted from parts and experienced wholly in time as a joined experience between audience and performer? Goehr has famously pointed out in \cite{Goehr1989} that the ontological status of that which we call a `musical work' is ``an open concept with paradigmatic and derivative employment''. Being an \emph{open concept} (first described in \cite{Weitz_1972}) thereby grants musical works status as a challenge to the universality of OOO when it comes to the arts - how can we rigidly define what a musical work is in terms of, say ANT, when it is perfectly valid to change the nature of that object-relation definition at any level and still have a valid musical work? From the requirements to be verified as such, \emph{e.g.} players, a venue, an audience, a score, a recording, \emph{etc.}, or even down to the definition of the nature of the base objects considered, such as computer composed music, atypical musical instruments, or just plain silence - all of these parameters themselves have artistic interpretation.

Such questions leave the core of this work's intentions in a unique position as creating a challenge of a strong kind to the philosophical theories that do not admit entanglement; not just of creating, but directly imitating and channeling a phenomenon that is not stipulated to be possible.

\subsection{Entanglement in Musical Performance}\label{sec:jazzentanglement}

There are a growing number of examples of exploring these human-computer and human-object entanglements from the view of music, for example Mice and McPherson in \cite{Mice_and_McPherson_2023} present a fascinating analysis of how these entanglement theories influence the design and operation of large-scale digital music instruments and performances. 

This work builds on an analysis of how entanglement theories relate to the interaction between humans and their musical instruments in \cite{Waters2021}, in which the starting thesis may be one of the most tantalising research questions in experimental instrument development; to what part is a musical instrument found and to what extent is it designed or made?

There has also been much discussion about the way in which performers can experience `entanglement' during performance or in the context of inter-human synchronization, \emph{cf.} \cite{Bhave2023}, \cite{Gloor2022}, \cite{Yuan2023}. 

Gloor, Zylka, and Fronzetti in \cite{Gloor2022} define `entanglement' as a ``metric to measure how synchronized communication between team members is.'' They utilise a graph network to describe the `degree/betweenness centrality' of each agent, that is, the number of agents it is connected to in a communication network (degree) or the number of paths that pass through it (betweenness), and the shortest distance between two agents across the graph in order to determine how `entangled' both agents are. More specifically \cite{Gloor2022} defines the entanglement formula, which maybe be simplified for this discussion as \begin{align} E(x,y) = \frac{C_D(x) C_D(y)}{d(x,y)}\end{align} where $C_D(x)$ is the degree of $x$ in the network graph, and $d(x,y)$ is the shortest distance between $x$ and $y$ by number of edges traversed travelling from $x$ to $y$ across the network. 

There are a number of ways the author can see to improve this, \emph{e.g.} by particularly developing the use of Betweenness Centrality (see \cite{Prountzos2013}), as opposed to Degree Centrality, as such a change would give a much richer picture around the true interconnectedness across a network. However, it is particularly interesting to see a formula from a very different area of science which approaches the same idea in a very different way.

Whilst \cite{Yuan2023} takes a very systematic view of inter-human connectedness, it is an interesting discussion around how to overcome some of the issues mentioned earlier with ANT. Yuan achieves this by embedding the epistemological chain in the knowledge creation process within the ANT network itself. Thereby, human-object interactions are represented by interactions along knowledge management within a given organization. Whilst this author is not entirely convinced that this resolves the deeper issues, it is certainly enough to allow a statement that actors interacting across a network can indeed be said to be meaningfully `entangled'. 

Bhave, van Delden, Gloor, and Renold bring entanglement and the study of human interconnectedness to a thoroughly musical analysis. Their findings are very interesting, and may be summarised in two main points quoted from \cite{Bhave2023}:

\begin{enumerate}
    \item ``Musicians are more likely to be entangled when they collectively practice more to attain perfect synchronization.''
    \item ``Synchronization in head, arm, and leg movements among musicians indicates strong team entanglement and a state of flow with the musicians.''
\end{enumerate}

Whilst Bhave's focus is on jazz musicians in particular, this level of synchronicity between musicians is what this work intends to find ways to extract and then represent in the musicians' instruments. Effectively, making physical what is currently only ephemeral or metaphysical. Extending into the real what is to date a mostly human experience. 

This section hopes to have briefly illuminated the intersection of various definitions of entanglement that is the core focus of this work. Is it possible to mimic such human-to-object/human-to-human entanglement in a physical entanglement between two objects? 

\subsection{Motivation}

With this theoretical backdrop, the intention and motivation for this work is to attempt to create some meaningful simulacra of quantum entanglement within these more human-centric propositions. 

In line with the starting point in the abstract of \cite{Waters2021}, the aim is to provide tools for the discovery of musical expression and the shaping of musical intention by means of interaction directly or alongside quantum phenomena. Just as artists explore the intra-human musical space, could their instruments likewise be entangled in both musical and physical senses?

Of course, this is very difficult to do with `real' photons. The main starting point for this author's previous work is to make use of high-fidelity simulations of quantum effects that can be pushed into hardware to reduce latency. The work presented herein extends many notions first explored in \cite{Carney_2023}, with the aim to realise a research thread that was first proposed in that document - namely that of entangling instruments that in some way reflects the entanglement between the musicians playing them.

\section{Quantum Phenomena and their Simulation}\label{sec:qit}

This section will briefly state the technical minimum that is needed to describe fully the theoretical work actuated later in this chapter. Throughout this chapter, the main resource for technical aspects of Quantum Information Theory (QIT) is \cite{Nielsen_Chuang_2010}. 

\subsection{Background Quantum Information Theory}

There are only a few mathematical requirements to build entanglement. In QIT, all states are usually measured in the same basis, yielding all measurements to be 0 or 1. These quantum bits, or `qubits' for short, form the basis of quantum information theory. Note that, whilst there is a growing theory concerning alternatives, such as qudits in \cite{Krishna2016}, this chapter will focus on qubits and quantum circuits.

A qubit's state shall be denoted using $\ket{\psi} = \begin{psmallmatrix} a \\ b \end{psmallmatrix} = a\ket{0} + b\ket{1}$, where $a,b \in \mathbb{C}$ with the only requirement such that $|a|^2 + |b|^2 = 1$. By convention $\ket{0} = \begin{psmallmatrix} 1 \\ 0 \end{psmallmatrix}$ and $\ket{1} = \begin{psmallmatrix} 0 \\ 1 \end{psmallmatrix}$. Operations on single or multiple qubits take the form of unitary operator matrices. For single qubit gates these are $2 \times 2$ matrices, whilst other matrices are built from tensor products of these to form $2^n \times 2^n$ matrices for $n$-qubit gates. 

The only three quantum operators, also called `quantum gates', that are required for this work are the $X$ gate (also called the $NOT$ gate), Hadamard or $H$ gate, and Controlled-NOT or $CNOT$ gate denoted \begin{align}X = \begin{pmatrix} 0 & 1 \\ 1 & 0 \end{pmatrix} \\ H =  \frac{1}{\sqrt{2}} \begin{pmatrix}  1 & 1 \\ 1 & -1 \end{pmatrix} \\ CNOT = \begin{pmatrix}  1 & 0 & 0 & 0 \\ 0 & 1 & 0 & 0 \\ 0 & 0 & 0 & 1 \\ 0 & 0 & 1 & 0 \end{pmatrix}\end{align}

We can see why $X$ is called then $NOT$ gate as $X\ket{0} = \ket{1}$ and $X\ket{1} = \ket{0}$. Meanwhile the $H$ gate is usually utilised to place a qubit into `superposition' - a balanced state where measurement may result in \emph{either} a 0 or 1 reading, with equal 50\% of each occurring. A different notation for this is given by:

\begin{align}
    H\ket{0} = \frac{\ket{0}+\ket{1}}{\sqrt{2}}
\end{align}

The Controlled-NOT gate is a very interesting 2-qubit gate that applies an $X$ gate to the second qubit only if the first qubit is in state $\ket{1}$. The $CNOT$ truth table is given by:

\begin{align}
\begin{split}
CNOT\ket{00} = \ket{00}, & \text{    } CNOT\ket{01} = \ket{11} \\
CNOT\ket{10} = \ket{10}, & \text{    } CNOT\ket{11} = \ket{01}
\end{split}
\end{align}


\subsubsection{Quantum Circuits and Entanglement}

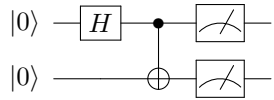
\begin{figure}[b]\label{fig:entanglement}
$$ \Qcircuit @C=1em @R=.7em {& \lstick{\ket{0}} & \gate{H} & \ctrl{1} & \meter & \qw \\ & \lstick{\ket{0}} & \qw & \targ & \meter & \qw} $$
\caption{A simple entanglement circuit.}
\end{figure}

In lieu of detailing matrices and tensor products, there is a simple graphical manner in which quantum algorithms are presented, known as `quantum circuits'. This graphical representation is an especially useful shorthand that has been extended into other graphical diagrammatic paradigms, such as category theoretic string diagrams in a language known as $ZX$-calculus \cite{Coecke2023}. That said, this chapter will stick to regular quantum circuit diagrams. Indeed, a simple quantum circuit that describes entanglement is given in figure \ref{fig:entanglement}.

Using this diagram, it can be shown that entanglement may arise from the following calculation: 

\begin{align}
\begin{split}
    CNOT(H\ket{0} \otimes \ket{0}) &= CNOT \frac{\ket{00}+\ket{01}}{\sqrt{2}} \\
    &= \frac{\ket{00}+\ket{11}}{\sqrt{2}}
\end{split}
\end{align}

Reading this equation it can be seen that the two outcomes for measurement will be 50\% `00' or 50\% `11', and that it is not possible to separate the two individual qubit states in this composed state. As such, this is the base entangled state that this work shall develop later for musical and artistic purposes.

Lastly, it should be noted that this is the first kind of entanglement. There is another option, which comes when you apply an $X$ gate to the second qubit. This changes the calculation to the following: 

\begin{align}
\begin{split}
    CNOT(H\ket{0} \otimes \ket{1}) &= CNOT \frac{\ket{10}+\ket{11}}{\sqrt{2}} \\
    &= \frac{\ket{10}+\ket{01}}{\sqrt{2}}
\end{split}
\end{align}

The probabilities for measurement will be 50\% `01' or 50\% `10', with the two qubits still being necessarily inseparable. This `$X$ switch' will be utilised later as a parameter for our simulations. Let the standard notation for these two states be given as 

\begin{align} \ket{\Phi^+} = \frac{\ket{00} + \ket{11}}{\sqrt{2}}, \ket{\Psi^+} = \frac{\ket{01}+\ket{10}}{\sqrt{2}} \end{align} These are widely known as the `Bell states' after the seminal paper due to Bell in \cite{Bell1964}. It should be noted that there are indeed $\ket{\Phi^-}$ and $\ket{\Psi^-}$ states, differing in phase from their $^+$ counterparts by replacing the plus signs in the numerators with minus signs. They are mentioned here only for completeness, and are not useful for the particular circuit in play later on. This is due to the fact that everything in this work ignores global phase, more details on how and why may be found in \cite{Carney_2023} and \cite{Nielsen_Chuang_2010}.

An important idea to note from some of the hype around entanglement is that when you separate and measure one qubit, you may already know the state of that qubit depending on the precise entanglement you have configured. But there is no way to communicate this faster than the speed of light. 

\section{Methodology}\label{sec:methodology}

This section details the algorithmic and processing approaches, as well as the necessary tooling to realize these ideas musically.

\subsection{Musical Interpretation of Entanglement}

To present this methodology, it is important to first define the circuit that will be used in our simulations, and then the way in which the musicians' inputs will be interpreted as well as the way that the entanglement will be distributed in these simulations. 

\textbf{Core Idea} - if the players' are musically closer to each other, defined here as taking the average tone value for each player, their instruments are entangled in $\ket{\Phi^+}$ type entanglement. This means that they will share precisely the same random values from the quantum simulation. As they move musically apart the simulations will go from $\ket{\Phi^+}$, through balanced random output, and eventually to $\ket{\Psi^+}$ entanglement, whereupon the players' instruments will have precisely opposite configurations. This is how this work realises the core artistic intention from the opening of this chapter.

\subsubsection{Parameterized Entanglement Swtich Circuit}

First add one extra gate to the QIT toolbox, the $R_x(\theta)$ gate defined from \cite{Nielsen_Chuang_2010} as:

\begin{align}
    R_x(\theta) = \begin{pmatrix} \cos(\theta/2) & -i \sin(\theta/2) \\ -i \sin(\theta/2) & \cos(\theta/2) \end{pmatrix}
\end{align}

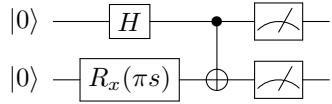
\begin{figure}[t]
    \centering
    $$ \Qcircuit @C=1em @R=.7em {& \lstick{\ket{0}} & \gate{H} & \ctrl{1} & \meter & \qw \\ & \lstick{\ket{0}} & \gate{R_x(\pi s)} & \targ & \meter & \qw} $$
    \caption{The main entanglement switching circuit with parameter $0 \leq s \leq 1$.}
    \label{fig:eswitch}
\end{figure}

This permits the usage of the same circuit the author has used previously in \cite{Carney_2023}, which may be found in figure \ref{fig:eswitch}.

This circuit allows a parameter to be used to `switch' between the two main entanglement states by means of the $R_x$ rotation gate. To learn more about why these are rotations, as well as discussions around the Bloch sphere as a visual aid and intuition pump for QIT, see \cite{Nielsen_Chuang_2010}. The main vector representation of the primary operation of this circuit, known as the statevector \cite{Nielsen_Chuang_2010}, for this circuit may be given by 

\begin{align}\label{eqn:state} \big( \cos(s/2)\ket{\Phi^+} + \sin(s/2)\ket{\Psi^+} \big) \end{align} 

Note that as the parameter $s$ sweeps from 0 to 1, at $s=0.5$ the circuit will output a balanced superposition of the form 

\begin{align}
    \frac{\ket{00}+\ket{01}+\ket{10}+\ket{11}}{\sqrt{4}}
\end{align} 

This gives the possibility of having a continuous spectrum of possible outputs. It is these outputs that the $R_x$ gate in figure \ref{fig:entanglement} allows us to scale between - after all, a switch in the quantum world has a continuum of possible superpositions of its two output values!

\subsubsection{Parameter Extraction}

In prior work \cite{Carney_2023} the author mapped the MIDI spectrum to $1/64$ increments of the $s$ parameter. However, in this work the intention is more subtle and so the mapping chosen is restricted to $1/12$ increments, effectively mapping the octave to the half-rotation of a qubit in semitone increments of $\pi/12$.

To explain why, first define a `tonal similarity' to be the distance between the average values for two arrays of note values that have been played so far in a given performance. In a naive way, this expresses the similarity, and thereby synchronicity, between two musicians as the distance between the notes they are playing. By reading the values of players' notes from MIDI it is easy to calculate this. 

This tonal similarity is generated by taking the average of a fixed window of notes played and sampled by the MIDI input from two players (in this case, the previous 8 notes), and then comparing the distance between these averages. This is then scaled to fit in the parameter $0 \leq s \leq 1$ and passed to the circuit for simulation. Of course, there are many more ways to perform this `musical centrality' analysis, which we will discuss later. 

The difference $\delta$ between these averages is taken as $(\delta \mod 12)$, and then scaled to between $0 \leq \delta \leq 1$. This scales a tonal center of 1 octave either side (positive or negative) from the players' tonal centres to gauge their tonal similarity in a fast, efficient, if slightly naive way. 

\subsubsection{Entangled Effect Output Generation}

With the parameter taken from the tonal similarity outlined above, the circuit is prepared and simulated multiple times. The output bit pairs from the simulations are then separated and sent to the instruments using MIDI Control Change (CC) Messages (see \cite{MIDI2014}) defined for each instrument separately. 

The intended effect of this is to actualize the difference in entanglement to a given fixed parameter. For example, on instrument may receive the entangled randomness to its second oscillator frequency, whilst the other to its main filter resonance setting. These are but two examples from a wide range of combinatorial possibilities. 

This is designed to mimic the way in which a physical entanglement system might work. For example, if one was to have access to an entangled photon generator, two fiber optic lines, and two photon detectors that can output `0' and `1' depending on the state of the photon they receive distributed from the source. By separating the bits that are simulated classically and passing them individually to each instrument over MIDI CC messages, we are simulating the distribution of entangled photons, as if the measurement of each had been performed ahead of transmission. 

\subsubsection{Intended Effects}

Whilst the output of the simulator is of course totally random, as would be the output from real quantum hardware, it is the manner in which the two output streams of random bits relate to one another that creates the interesting effects to be harnessed artistically. 

As a pair of players drift away from each other in reference to a tonal centre, so do the random streams of bits they receive over MIDI to their instruments. Going from being concordant in a $\ket{\Phi^+}$ state, with each getting matching values. If they play at octave average difference then they will receive bit-wise opposite values which will mean that one instrument will be boosted at random from the other. If they are playing at a 7-semitone tonal average difference then they will receive totally random values each - in many representing their respective drift from what is conventionally called `tonality'. 

It is these effects that the author intends to commit to artistic exploration by means of the devices and code constructed in this work and made available to others to experiment with.

\section{Tooling Quantum Entangled Instruments with Embedded Simulations}\label{sec:tooling}

This section explores the way in which the above algorithms are coded and the physical tooling requirements for creating this simulated entanglement. 

\subsection{Embedded Simulators}

To facilitate the above algorithms and methods this work augments the electronics found in \cite{Carney_2023}, specifically the electronics used in the initial quantum MIDI experiments. 

The same code from the \verb|micro-quantum| project in \cite{Carney_2022}, as utilised in \cite{Carney_2023}, is utilized here. The code for this project has been mostly extended from the original code used in \cite{Carney_2023}. 

There are other ways in which quantum simulators may be embedded aside from the tooling that is proposed below, which will be discussed in section \ref{sec:future}. This proof-of-concept hardware implementation is presented to demonstrate effectively and definitively that inter-instrument entanglement can indeed be simulated. 

\subsection{Tooling}

\begin{sidewaysfigure}
    \centering
    \includegraphics[width=\textwidth]{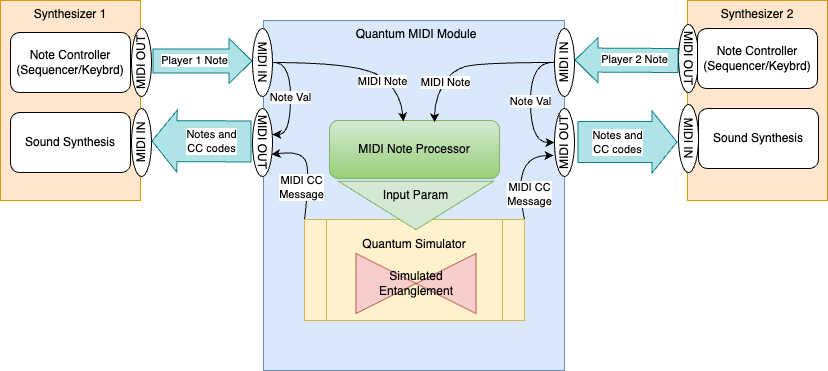}
    \caption{A block diagram of the MIDI Quantum Entanglement Simulator described in this chapter.}
    \label{fig:block-diag}
\end{sidewaysfigure}

A block diagram of the system that has been implemented in hardware may be found in figure \ref{fig:block-diag}. The Bill of Materials comprises the following items:

\begin{itemize}
    \item 1x Raspberry Pi Pico board
    \item 2x MIDI Breakout boards
    \item 4x MIDI cables
    \item An electronics breadboard and compatible jumper cables
    \item 2x MIDI Control Change message capable synthesizers
\end{itemize}

The precise synthesizers used for these initial experiments are a Yamaha Reface DX and a KORG NuTEKT NTS-1 mini-synth. Both of these can receive MIDI messages, however to increase the range of the KORG a Roland MIDI Controller Keyboard was used. 

The firmware on the Pico uses two separate interrupt driven UART peripherals (built into the chip) to handle the MIDI input and output. They are clocked at 31250Hz and transmit in bursts of three byte messages, as per the MIDI standard \cite{MIDI2014}.

The overall MIDI topology is thus; the central Q-MIDI device takes two inputs, one from the Yamaha and one from the Roland keyboard, and then passes the notes from the Yamaha MIDI OUT back to the Yamaha MIDI IN, whilst the notes from the Roland MIDI OUT are sent directly (and immediately so as to not introduce lag) to the KORG's MIDI IN. The simulation outputs are sent asynchronously from the input notes using the MIDI OUT ports from the simulator. 

The quantum simulations and average calculations are triggered every 100ms, so as to not inundate the MIDI CC messages and to let the settings be realized musically. The logic for commencing a quantum simulation is triggered by a new note being played by either player. 

The use of an embedded system reduces the latency for the MIDI note message relay to $< 1$ms which is well within the bounds indicated by \cite{McPherson_2020}. The fact that the simulations are performed asynchronously means that the musical effects to not perturb the musical expression from the players. 

\subsection{Entangled Synthesizer Parameters}

The entangled output bits are mapped to expressive synth parameters.

In order to be expressive, attention has to then be given to the various parameters that may be manipulated using MIDI CC messages, identified in each synthesizer's documentation.

To this end, the two parameters selected for each instrument were as follows - for the KORG, the output from the quantum simulator was taken as the output of the second qubit, and was sent to the main Oscillator Shape parameter. For the Yamaha, the output of the first qubit was extracted and sent to synthesizer as the Oscillator 2 Level.

In practical terms, this allows the players to hear how the system is interpreting their tonal similarity by means of the differences or synchronicity of their instruments. The more the musicians play the same note, the more their instruments will be shaped simultaneously in the same way. Meanwhile, if their tonal similarity differs, then the instruments will take opposite values for each parameter, which leaves it open as to how close they will realistically sound. 

This ability to directly affect timbre and sonority of a given instrument is the main reason that MIDI was chosen as the vehicle for the input/output mappings. Given most instruments can generate and/or understand MIDI, this code can be easily adapted to any MIDI-capable synthesizer of choosing.

\section{Results}\label{sec:results}

\begin{figure}[t]
    \centering
    \includegraphics[width=\textwidth]{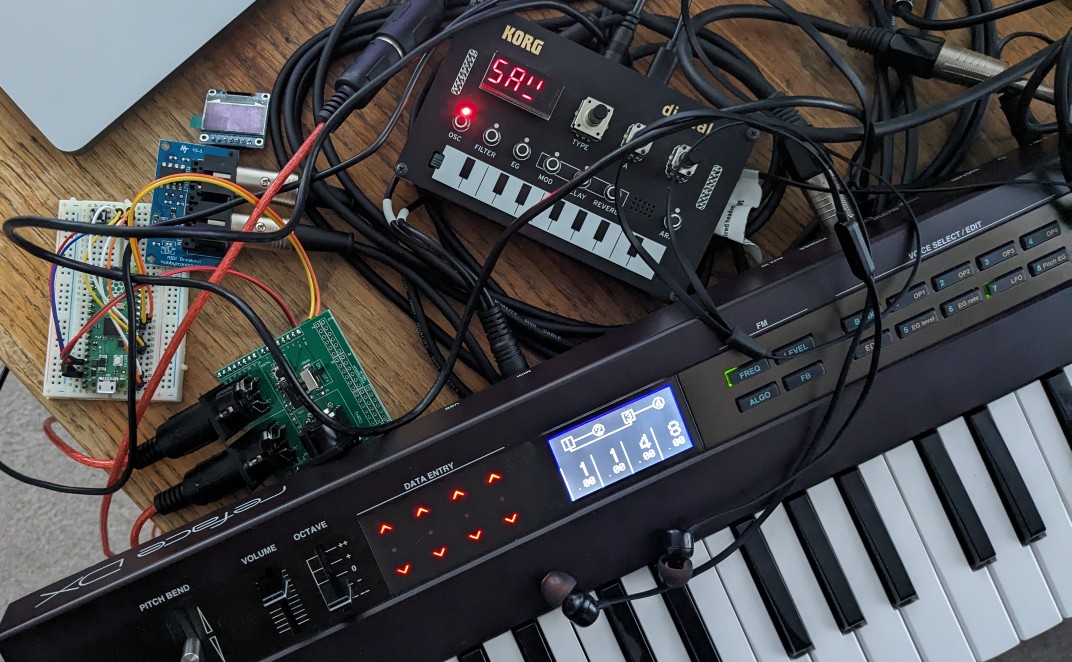}
    \caption{The experimental MIDI Quantum Entanglement setup, with the Yamaha Reface DX and KORG NTS-1 visible, connected to the simulator (left of picture) with two MIDI IN/OUT interface pairs. The Roland controller MIDI keyboard is out of frame.}
    \label{fig:setup}
\end{figure}

An image depicting the experiments and development setup for this proof of concept may be found in figure \ref{fig:setup}. An experimental improvisation may be found on the SoundCloud platform at \cite{Carney_2024_SoundCloud}. The source code and schematics for the electronics has been uploaded and added to the `Quantum Music \@ Quantum Village' repository on GitHub, which may be found at \cite{Carney_2022}.

The improvisation performance shows how the tonal centre - here defined by the persistence of the KORG drone - and its relationship to the Yamaha lead affects the way that the sounds of both instruments are manipulated. There are moments, for example near the start, where the Yamaha is concordant and tonally close to the KORG, whereupon the effects are heard clearly to be simultaneous and coincident. Meanwhile, when the lead Yamaha line drifts atonally away in a modal style the two synthesizers cease to have any likeness in how their parameters are manipulated; given the two lines have drifted, so have their parameters. 

Although this demonstration is short, it shows that the equipment works, and is viable for deeper and further development and experimentation. It certainly shows that the idea has validity and may be considered to for future developments within quantum computer music, many of which are discussed in section \ref{sec:future}.

From an artistic interface latency point of view, the author can attest that the interaction with the instruments was no different with or without the quantum simulator. There was quite a lot of debugging required to make this true for both instruments, however - the KORG has some issues with fast slews of MIDI messages owing to its own fairly limited resources. The Yamaha, with a more powerful processor, had no issues. With minimal tweaking, it was possible to play both instruments at the same time and experience the tonal shifts based on the quantum simulator's `entangled interpretation' of what was being played.

\subsection{Practical Considerations}

The only practical limits of this system was from the use of local electronics and fixed-length MIDI cables. If there were a way to extend these - physically or virtually - then there is no bound on where two musicians may interact with this simulated entanglement. 

Of course, the real aim will be to use real hardware, of which there are known limits with fibre optics, for example. This will be discussed later in more detail in section \ref{sec:future}.

The latency from the note relay over MIDI was imperceptible, and the transition timings from current parameter values to new parameter values and the interval between simulation events were fine tuned to compliment a variety of musical tempi and harmonic rhythms. These parameters were `5\% increments per 10ms' transition speed, and 100ms between simulations. These are found in the source code in the GitHub repository. 

It is worth noting that there is no necessary requirement for the specific synthesizers that have been chosen, mainly due to availability and ease of programming, in this work. Editing the MIDI code in \cite{Carney_2022} to suit any MIDI CC configurable synthesizer is relatively straightforward, and may be done so by editing the parameters \verb|P1_CTRL_CODE| and \verb|P2_CTRL_CODE| in the \verb|qmidi.c| file. 

\section{Discussion}\label{sec:disc}

This work has demonstrated two key ideas; firstly, it is indeed possible to simulate quantum entanglement between two instruments, and even to go as far as make such a simulation reactive to musical performance. Secondly, there is a need for discussing and interpreting the entanglement between musicians, their instruments, the audience, and other musicians.

There are, of course, many more questions that lie open. For instance, is the `tonal centrality' algorithm described and instantiated in hardware optimal? This author thinks not - it is deliberately naive so that it may be eliminated as a point of error or failure during development of the overall experimental setup. 

The feeling of playing with another musician was not palpably different for this experimental setup, however there was definitely a sense that there was more reactivity from the instruments the author was playing. The sonic tension from the instruments in agreement or set contrariwise against each other is sonically interesting. But is it musically interesting? 

There are, of course, many interesting questions that arise from the possibility of a `physically entangled' pair of instruments. For example, who is really doing the observing in this situation - the simulator, the musicians, or the audience? 

What can definitely be said about the setup described above is that there is now another way to add `quantum colour' to musical instruments, and in a way that meaningfully connects two instruments whilst setting them in relation to the connection between the players. The author believes that this is the most necessary exploration of the quantum musical space - to find new ways for musicians to shape sound with quantum, as they sculpt time with sound. 

\section{Future Work}\label{sec:future}

There are many avenues for which this work may be developed and carried forward, some of the more interesting of which we discuss below.

\subsection{Real Entanglement}

Of course, with a simulated prototype now existing, the next question is whether this can be done with \emph{real} quantum hardware. This may look something like the following: two instruments each connected to a photon detector that are coupled to the output of an entanglement source. These detectors report on a given measurement axis, such as polarization. These are then connected to two outputs of an entanglement source that generates pairs of photons. 

The real question then becomes this; what do we do to the entanglement source to interpret the sounds from the musicians to capture something of their musical entanglement? Recent advances in creating programmable optical circuits, such as the work by Malik and their team in \cite{Goel2024}, may well pave the way towards methods to enable this kind of feedback in real hardware. 

\subsection{Artistic Influence}

With experimental setups such as presented in this chapter, it does beg the question about whether there is any real musical merit in harnessing these quantum effects. 

Does such a notion of entanglement scale upwards, and if it does, is there any artistic merit in doing so? The real answers to this will only come about from development and dissemination of the equipment to create and work with, at the very least, simulated entanglement. To this end, the source code from this experiment is available free and open source for immediate experimentation in \cite{Carney_2022}.

\subsection{`Musicality Extraction' Methods}

The `tonal centrality' algorithm is by no means very nuanced. But what would be a better replacement? Would it be worthwhile setting not just a tonal centrality from an average of notes, but perhaps from scales themselves? Is it prudent to estimate to what extent a musician is playing diatonically? Or modally? And then map the difference in this between two musicians as a matter of dynamic measurement on their performance?

Taking physical measurement depicted in \cite{Bhave2023}, should this be used as a way to gauge entanglement for a pair or more of musicians? 

\subsection{Qudits}

With recent advances in quantum hardware it has been possible to create natively entangled \emph{qudits}, as recently demonstrated by Hrmo and their team at ETH in \cite{Hrmo2023}. The qudit - short for `quantum digit' - has had relatively little attention in the Quantum Music space, but it has heralded some significant improvements in the manipulation of the Hilbert space around a quantum state, see \cite{Fan2007}, \cite{Hunt2020}, and \cite{Krishna2016} for examples. 

Qudits have always been considered to have interesting entanglement potential, for example going back to \cite{Rungta2000}. This is something that this author considers an interesting way in which to entangle multiple `artistic parameters' into just a few entangled qudits.

\subsection{$>2$ Instrument Entanglement and GHZ States}

Entanglement in physics is not exclusive between only two particles. It has been known for some time that GHZ states of multiple entangled particles are computationally very powerful (\emph{e.g.} see \cite{D'Hondt2005}). Likewise musically, ensembles have always grown past just a duet. This leads a natural question to be can we simulate entanglement on more and more instruments? How does entangling instruments affect or represent the entanglement between musicians in a quartet? Or a Jazz Big Band? Or even an \emph{Entangled Orchestra}? 

\subsection{Entangled Compositions}

When considering more and more players, it can also arise the notion of what `entangled composition' may look like? The music of Morton Fledman or John Cage have famously incorporated the random into their music - the ``music of chance'' from the New York Modernists feature moments of total musical freedom, undefined formally, constrained only by what goes before and after \cite{Dohoney2017}. 

Of course, such randomness is intrinsic to the measurement of a quantum system, with quantum entanglement allowing us to choose in what manner that randomness manifests relationally - we may not know precisely what will be measured, but we know the relationship of this measurement between two or more parties. So what, then, does a `Quantum Entangled Symphony' involve? 

\section{Acknowledgements}

The author would like to thank the editors, in particular Prof. Miranda for the invitation to prepare this chapter. The author particularly thanks Victoria Kumaran, Peter Wolfendale, and Il\=a Kamalagharan for their indulgent conversations, references, and resources, as well as Prof. Ben Varcoe for his continued academic inspiration.

\bibliographystyle{ws-book-har}   
\bibliography{mc-bibliography}      









\printindex

\end{document}